\documentclass[pra,twocolumn,showpacs,preprintnumbers,amsmath,amssymb,aps]{revtex4}
\usepackage{color}
\usepackage{bm, graphicx, amsmath}
\usepackage{hyperref}

\usepackage{braket}

\begin{document}
\title{Average concurrence and entanglement swapping}
\author{J\'anos A. Bergou$^{1,2}$, Dov Fields$^{1,3}$, Mark Hillery$^{1,2}$,  Siddhartha Santra$^{3}$, Vladimir S. Malinovsky$^{3}$}
\affiliation{$^{1}$ Department of Physics and Astronomy, Hunter College of the City University of New York, 695 Park Avenue, New York, NY 10065 \\
$^{2}$Graduate Center of the City University of New York, 365 Fifth Avenue, New York, NY 10016 \\
$^{3}$United States Army Research Laboratory, Adelphi, Maryland 20783} 
\begin{abstract}
We study the role of average concurrence in entanglement swapping in quantum networks.  We begin with qubit pure states, and there is a very simple rule governing the propagation of average concurrence in multiple swaps.  We look at examples of mixed qubit states, and find the relation for pure states gives an upper bound on what is possible with mixed states.  We then move on to qudits, where we make use of the I-concurrence.  Here the situation is not as simple as for qubits, but in some cases relatively straightforward results can be obtained. 
\end{abstract}

\maketitle

\section{Introduction}
Entanglement swapping is a basic protocol in quantum information \cite{Zukowski,Zeilinger,Gisin}.    One starts with two pairs of entangled particles, then takes one particle from each pair, and finally measures them in an entangled basis.  The result is that the two remaining particles are now entangled, whereas they were not initially.  Entanglement swapping has many uses.  It can be used to create multipartite entangled states from singlets \cite{bose} and as a method of entanglement purification \cite{knight}.  It can serve as the basis for a quantum repeater, which would enable the transmission of entanglement over long distances \cite{briegel,duan,razavi,lukin}.

Here we wish to look at the connection between entanglement swapping and concurrence.  In particular, we want to study how the concurrences of the input states are related to the average coherence of the output state.  We first summarize the results for qubit pure states, and discuss earlier results \cite{knight,network}.  For pure states, the average concurrence of the final state is found to be just the product of the concurrences of the states used to do the entanglement swapping.  This result generalizes to multiple entanglement swaps.  We then go on to look at concurrence and entanglement swapping for noisy qubits.  Entanglement swapping of noisy qubits has been studied in connection with Bell nonlocality \cite{sen,wojcik,klobus}, enhancing singlet fraction \cite{grudka}, and for $X$ states \cite{roa}.  An extensive study of the concurrence that can be obtained by the entanglement swapping of noisy qubits was done in \cite{kirby}, and what we do here can be viewed as complementary to that work.  There the probabilities to find different values of the concurrence were not incorporated, while here, since we are examining average concurrence, they are.  We then proceed to examine entanglement swapping for qudits \cite{hardy,bouda}, and we use I-concurrence for our concurrence measure \cite{buzek}.  Qudits can be used in quantum networks, and recent work suggests that qudit networks can be realized and have some advantages \cite{bacco}.  We conclude with a study of I-concurrence and entanglement swapping of noisy qudit states.

\section{Entanglement swapping of qubits}

\subsection{Pure states}
We will start with two two-qubit entangled states,  one being a state of qubits $a$ and $b$, and the other being as state of qubits $c$ and $d$.  These states can be expressed in terms of their Schmidt bases, so we have
\begin{eqnarray}
|\psi\rangle_{ab} & = & \sqrt{\lambda_{0}} |u_{0}\rangle _{a}|v_{0}\rangle_{b} + \sqrt{\lambda_{1}} |u_{1}\rangle_{a}|v_{1}\rangle_{b} \nonumber  \\
|\psi\rangle_{cd} & = & \sqrt{\lambda_{0}^{\prime}} |u_{0}^{\prime}\rangle _{c}|v_{0}^{\prime}\rangle_{d} + \sqrt{\lambda_{1}^{\prime}} |u_{1}^{\prime}\rangle_{c}|v_{1}^{\prime}\rangle_{d} .
\end{eqnarray}
Here $\lambda_{0}+\lambda_{1}=1$ and $\langle u_{j}|u_{k}\rangle = \langle v_{j}|v_{k}\rangle = \delta_{jk}$, and similar conditions hold for the primed quantities.  The initial state for the procedure is
\begin{equation}
|\Lambda\rangle  = |\psi\rangle_{ab} |\psi\rangle_{cd} .
\end{equation}

Defining the measurement basis for the $bc$ qubits
\begin{eqnarray}
|\tilde{\Psi}_{+}\rangle & = & \alpha_{0}|v_{0}\rangle_{b}|u_{0}^{\prime}\rangle_{c} + \beta_{0} |v_{1}\rangle_{b}|u_{1}^{\prime}\rangle_{c} \nonumber \\
|\tilde{\Psi}_{-}\rangle & = & \beta_{0}^{\ast}|v_{0}\rangle_{b}|u_{0}^{\prime}\rangle_{c} - \alpha_{0}^{\ast} |v_{1}\rangle_{b}|u_{1}^{\prime}\rangle_{c} \nonumber \\
|\tilde{\Phi}_{+}\rangle & = & \alpha_{1} |v_{0}\rangle_{b}|u_{1}^{\prime}\rangle_{c} + \beta_{1}|v_{1}\rangle_{b}|u_{0}^{\prime}\rangle_{c} \nonumber \\
|\tilde{\Phi}_{-}\rangle & = & \beta_{1}^{\ast} |v_{0}\rangle_{b}|u_{1}^{\prime}\rangle_{c} - \alpha_{1}^{\ast} |v_{1}\rangle_{b}|u_{0}^{\prime}\rangle_{c} ,
\end{eqnarray}
where $\alpha_{j}$ and $\beta_{j}$ remain to be determined, and $|\alpha_{j}|^{2}+|\beta_{j}|^{2}=1$ for $j=0,1$. We then have that 
\begin{eqnarray}
|v_{0}\rangle |u_{0}^{\prime}\rangle & = &  \alpha_{0}^{\ast} |\tilde{\Psi}_{+}\rangle + \beta_{0} |\tilde{\Psi}_{-}\rangle \nonumber \\
|v_{1}\rangle |u_{1}^{\prime}\rangle & = & \beta_{0}^{\ast} |\tilde{\Psi}_{+}\rangle - \alpha_{0} |\tilde{\Psi}_{-}\rangle \nonumber \\
|v_{0}\rangle |u_{1}^{\prime}\rangle & = & \alpha_{1}^{\ast} |\tilde{\Phi}_{+}\rangle + \beta_{1} |\tilde{\Phi}_{-}\rangle \nonumber \\
|v_{1}\rangle |u_{0}^{\prime}\rangle & = & \beta_{1}^{\ast} |\tilde{\Phi}_{+}\rangle - \alpha_{1} |\tilde{\Phi}_{-}\rangle .
\end{eqnarray}
This then gives us for $|\Lambda\rangle$ 
\begin{eqnarray}
\label{lambda}
|\Lambda\rangle & = & (\sqrt{\lambda_{0}\lambda_{0}^{\prime}}\alpha_{0}^{\ast} |u_{0}\rangle_{a}|v_{0}^{\prime}\rangle_{d} + \sqrt{\lambda_{1}\lambda_{1}^{\prime}}\beta_{0}^{\ast} |u_{1}\rangle_{a}|v_{1}^{\prime}\rangle_{d}) |\tilde{\Psi}_{+}\rangle_{bc} \nonumber \\
&&+ (\sqrt{\lambda_{0}\lambda_{0}^{\prime}}\beta_{0} |u_{0}\rangle_{a}|v_{0}^{\prime}\rangle_{d} - \sqrt{\lambda_{1}\lambda_{1}^{\prime}} \alpha_{0} |u_{1}\rangle_{a}|v_{1}^{\prime}\rangle_{d}) |\tilde{\Psi}_{-}\rangle_{bc} \nonumber \\
&&+ ( \sqrt{\lambda_{0}\lambda_{1}^{\prime}} \alpha_{1}^{\ast} |u_{0}\rangle_{a}|v_{1}^{\prime}\rangle_{d} + \sqrt{\lambda_{0}^{\prime}\lambda_{1}}\beta_{1}^{\ast} |u_{1}\rangle_{a}|v_{0}^{\prime}\rangle_{d}) |\tilde{\Phi}_{+}\rangle_{bc} \nonumber \\
&& + (\beta_{1} \sqrt{\lambda_{0}\lambda_{1}^{\prime}} |u_{0}\rangle_{a}|v_{1}^{\prime}\rangle_{d} - \sqrt{\lambda_{0}^{\prime}\lambda_{1}} \alpha_{1} |u_{1}\rangle_{a}|v_{0}^{\prime}\rangle_{d}) |\tilde{\Phi}_{-}\rangle_{bc} ]  . \nonumber \\
\end{eqnarray}
In order to compute the average concurrence of the $ad$ state after measuring in the above $bc$ basis, we multiply the probability of getting a particular $bc$ state times the concurrence of the resulting $ad$ state.  For example, we obtain the result $|\tilde{\Psi}_{+}\rangle_{bc}$ with probability 
\begin{equation}
P_{+}= \lambda_{0} \lambda_{0}^{\prime} |\alpha_{0}|^{2} + \lambda_{1} \lambda_{1}^{\prime} |\beta_{0}|^{2} ,
\end{equation} 
and the concurrence of the resulting state is $C_{+}=2(\lambda_{0}\lambda_{0}^{\prime}\lambda_{1}\lambda_{1}^{\prime})^{1/2} |\alpha_{0}\beta_{0}| / P_{+}$, so the contribution of this measurement result to the average concurrence is $P_{+}C_{+}$.  Adding up all of the contributions we find that
\begin{equation}
C_{av}= 4 ( |\alpha_{0}\beta_{0}| + |\alpha_{1}\beta_{1}| ) \sqrt{\lambda_{0}\lambda_{1}\lambda_{0}^{\prime}\lambda_{1}^{\prime}} .
\label{avconc_maxbasis}
\end{equation}
This is maximized when $\alpha_{j}=\beta_{j} = 1/\sqrt{2}$, for $j=0,1$.  In that case we have that \cite{network}
\begin{equation}
\label{out1}
C_{av}=4 \sqrt{\lambda_{0}\lambda_{1}\lambda_{0}^{\prime}\lambda_{1}^{\prime}} .
\end{equation}
Noting that the concurrences of the two input states are $C_{ab}= 2\sqrt{\lambda_{0}\lambda_{1}}$ and $C_{cd} = 2 \sqrt{\lambda_{0}^{\prime}\lambda_{1}^{\prime}}$, we find
\begin{equation}
C_{av}=C_{ab}C_{cd} .
\end{equation}
This is less than or equal to either $C_{ab}$ or $C_{cd}$ unless one of the states is maximally entangled, so, in most cases, the use of non-maximally entangled states in entanglement swapping will degrade the average concurrence. 

We can next ask what happens if we do a second entanglement swap \cite{network}.  We start with Eq.\ (\ref{lambda}) and set $\alpha_{k}$ and $\beta_{k}$ equal to $1/\sqrt{2}$, for $k=0,1$.  We will denote the states in the measurement basis with this choice as before, but without the tildes.  If we measure the $bc$ qubits, the probability of obtaining $|\Psi_{\pm}\rangle_{bc}$ is 
\begin{equation}
P_\Psi  = \frac{1}{2}(\lambda_{0}\lambda_{0}^{\prime} + \lambda_{1}\lambda_{1}^{\prime}) ,
\end{equation}
and the probability of measuring either $|\Phi_{\pm}\rangle_{bc}$ is
\begin{equation}
P_\Phi = \frac{1}{2}(\lambda_{0}\lambda_{1}^{\prime} + \lambda_{1}\lambda_{0}^{\prime}) .
\end{equation}
If we measure either $|\Psi_{\pm}\rangle_{bc}$, then we append the state 
\begin{equation}
|\psi^{\prime\prime}_{1}\rangle_{ef} = \sqrt{\lambda_{0}^{\prime\prime}} |u_{0}\rangle_{e}|v_{0}^{\prime}\rangle_{f} + \sqrt{\lambda_{1}^{\prime\prime}} |u_{1}\rangle_{e}|v_{1}^{\prime}\rangle_{f} ,
\end{equation}
and if we measure either $|\Phi_{\pm}\rangle_{bc}$, then we append the state 
\begin{equation}
|\psi^{\prime\prime}_{2}\rangle_{ef} = \sqrt{\lambda_{0}^{\prime\prime}} |u_{0}\rangle_{e}|v_{1}^{\prime}\rangle_{f} + \sqrt{\lambda_{1}^{\prime\prime}} |u_{1}\rangle_{e}|v_{0}^{\prime}\rangle_{f} .
\end{equation}
We then measure the $de$ qubits in the Bell basis.  If we found  $|\Psi_{\pm}\rangle_{bc}$ in the first measurement, then then the average concurrence of the $af$ state is
\begin{eqnarray}
C_{av}^{\Psi} & = & \frac{1}{4 P_\Psi} 2\sqrt{\lambda_{0}\lambda_{1}\lambda_{0}^{\prime}\lambda_{1}^{\prime}} 2\sqrt{\lambda_{0}^{\prime\prime}\lambda_{1}^{\prime\prime}} \nonumber \\
 & = & \frac{1}{4 P_\Psi} C_{ab}C_{cd} C_{ef} ,
\end{eqnarray}
where $C_{ef} = 2\sqrt{\lambda_{0}^{\prime\prime}\lambda_{1}^{\prime\prime}}$ is the concurrence of $|\psi^{\prime\prime}_{1}\rangle_{ef}$.  Similarly, if we found  $|\Phi_{\pm}\rangle_{bc}$ in the first measurement, then then the average concurrence of the $af$ state is
\begin{eqnarray}
C_{av}^{\Phi} & = & \frac{1}{2 P_\Phi } 2\sqrt{\lambda_{0}\lambda_{1}\lambda_{0}^{\prime}\lambda_{1}^{\prime}} 2\sqrt{\lambda_{0}^{\prime\prime}\lambda_{1}^{\prime\prime}} \nonumber \\
 & = & \frac{1}{4 P_\Phi } C_{ab}C_{cd} C_{ef} ,
\end{eqnarray}
where $C_{ef} = 2\sqrt{\lambda_{0}^{\prime\prime}\lambda_{1}^{\prime\prime}}$ is the concurrence of $|\psi^{\prime\prime}_{2}\rangle_{ef}$, which is the same as that of $|\psi^{\prime\prime}_{1}\rangle_{ef}$.  Finally, the total average concurrence, $C_{av}$, is
\begin{equation}
C_{av} = 2( P_\Psi  C_{av}^{\Psi} + P_\Phi  C_{av}^{\Phi} ) = C_{ab}C_{cd} C_{ef} .
\end{equation}
This implies that if we do a string of entanglement swaps, the average concurrence between the first and the last qubit is just the product of the concurrences of the states we used to do the swaps.

The same reasoning can be applied to a more general situation.  Suppose we have an $N$-qubit state, $|\Psi_{N}\rangle$, and we single out one of the qubits, which we shall denote by $a$.  We have from the Schmidt representation that
\begin{equation}
|\Psi_{N}\rangle = \sum_{j=0,1} \sqrt{\lambda_{j}^{\prime}} |u_{j}\rangle_{a} |V_{j}\rangle_{\bar{a}} ,
\end{equation}
where $\,_{a}\langle u_{0}|u_{1}\rangle_{a} = \,_{\bar{a}}\langle V_{0}|V_{1}\rangle_{\bar{a}} = 0$, and the states $|u_{j}\rangle_{a}$ are single qubit states and $|V_{j}\rangle_{\bar{a}}$ are states of the $N-1$ qubits that are not $a$.  If we now append  the state $(1/\sqrt{2})\sum_{j=0,1} \sqrt{\lambda_{j}} |u_{j}\rangle_{c}|u_{j}\rangle_{b}$ to $|\Psi_{N}\rangle$, and measure qubits $a$ and $b$ in the Bell basis, then the average concurrence between qubit $c$ and the $N-1$ qubits $\bar{a}$ is $C_{bc}C_{a\bar{a}}$, where $C_{a\bar{a}} = 2\sqrt{\lambda_{0}^{\prime}\lambda_{1}^{\prime}}$ is the original concurrence between qubit $a$ and the $N-1$ qubits $\bar{a}$.  This relation also holds if the $N$ qubit state is the result of a previous entanglement swap, whose average concurrence is $C_{av}^{a\bar{a}}$.  Then the average concurrence after the entanglement swap with the $bc$ qubits is $C_{av}^{a\bar{a}} C_{bc}$.

\subsection{GHZ measurements}
We will start this section with a note on notation.  For the rest of the paper, we will take advantage of the fact that the Schmidt bases of the two parts of a bipartite state can be transformed into computational bases by local unitary transformations.  Such transformations do not affect the entanglement, so we can assume that the Schmidt bases are the computational bases.  Now suppose we start with three entangled pairs and make a measurement in the GHZ basis in order to create a GHZ state \cite{bose,sen}.  In more detail, we start with the three states, in Schmidt form,
\begin{eqnarray}
|\psi\rangle_{ab} & = & \frac{1}{\sqrt{2}} (\sqrt{\lambda_{0}}  |00\rangle_{ab} + \sqrt{\lambda_{1}} |11\rangle_{ab} ) \nonumber \\
|\psi\rangle_{cd} & = & \frac{1}{\sqrt{2}} (\sqrt{\gamma_{0}}  |00\rangle_{ab} + \sqrt{\gamma_{1}} |11\rangle_{ab} ) \nonumber \\
|\psi\rangle_{ef} & = & \frac{1}{\sqrt{2}} ( \sqrt{\mu_{0}} |00\rangle_{ab} + \sqrt{\mu_{1}} |11\rangle_{ab} ) ,
\end{eqnarray}
and measure the $bdf$ qubits in the basis 
\begin{eqnarray}
|G_{0}^{\pm}\rangle & = & \frac{1}{\sqrt{2}} ( |000\rangle \pm |111\rangle ) \nonumber \\
|G_{1}^{\pm}\rangle & = & \frac{1}{\sqrt{2}} ( |001\rangle \pm |110\rangle )  \nonumber  \\
|G_{2}^{\pm}\rangle & = & \frac{1}{\sqrt{2}} ( |010\rangle \pm |101\rangle )  \nonumber  \\
|G_{1}^{\pm}\rangle & = & \frac{1}{\sqrt{2}} ( |100\rangle \pm |011\rangle )  .
\end{eqnarray}
The result will be a generalized GHZ state for the qubits $ace$.  For example, if we get $|G_{0}^{+}\rangle$ the resulting state is
\begin{equation}
\frac{1}{\sqrt{2p(G_{0}^{+})}} ( \sqrt{\lambda_{0}\gamma_{0}\mu_{0}} |000\rangle_{ace} + \sqrt{\lambda_{1}\gamma_{1}\mu_{1}} |111\rangle_{ace} ) ,
\end{equation}
where $p(G_{0}^{+}) = ( \lambda_{0}\gamma_{0}\mu_{0} + \lambda_{1}\gamma_{1}\mu_{1} )/2$ is the probability of obtaining $|G_{0}^{+}\rangle$.  

We can now view the resulting state as a bipartite state by singling out one qubit and looking at its entanglement with the other two.  In the example above, we could look at the entanglement between qubit $a$ and qubits $bc$.  This allows us to compute a concurrence, which we shall denote as $C_{21}$, and this concurrence does not depend on which qubit was singled out.  The average value of this concurrence  is $C_{21}^{(av)}$, and it is given by
\begin{equation}
C_{21}^{(av)} = C_{ab} C_{cd} C_{ef} ,
\end{equation}
so we again find a product rule for the final average concurrence.  This result also holds if the two-qubit concurrences are average concurrences.

\subsection{Noisy qubits}
We now want to look at entanglement swapping with qubits in states of the form
\begin{equation}
\label{noisyqubit}
\rho = p\frac{I}{4} + (1-p) |\psi\rangle\langle\psi |,
\end{equation}
where $|\psi\rangle = \sqrt{\lambda_{0}} |00\rangle + \sqrt{\lambda_{1}} |11\rangle$. These states are obtained due to partial depolarization of pure states $|\psi\rangle\langle\psi |$ through mixing with the completely depolarized state $I/4$ with the mixing parameter $p$. The form of the generalised Bell state here does not affect the essence of our conclusions; only its Schmidt coefficients are important.  The concurrence of these states is easy to evaluate if we consider their X-state \cite{xstate_yu} form, with nontrivial entries only along the diagonal and anti-diagonal, as evident in its matrix representation,
\begin{align}
\rho=
\begin{pmatrix}
\frac{p}{4}+(1-p)\lambda_0 && 0 && 0 && (1-p)\sqrt{\lambda_0\lambda_1}\\
0 && \frac{p}{4} && 0 && 0\\
0 && 0 && \frac{p}{4} && 0\\
(1-p)\sqrt{\lambda_0\lambda_1} && 0 && 0 && \frac{p}{4}+(1-p)\lambda_1
\end{pmatrix} .
\end{align}

\begin{figure}
\centering
\includegraphics[width=\columnwidth]{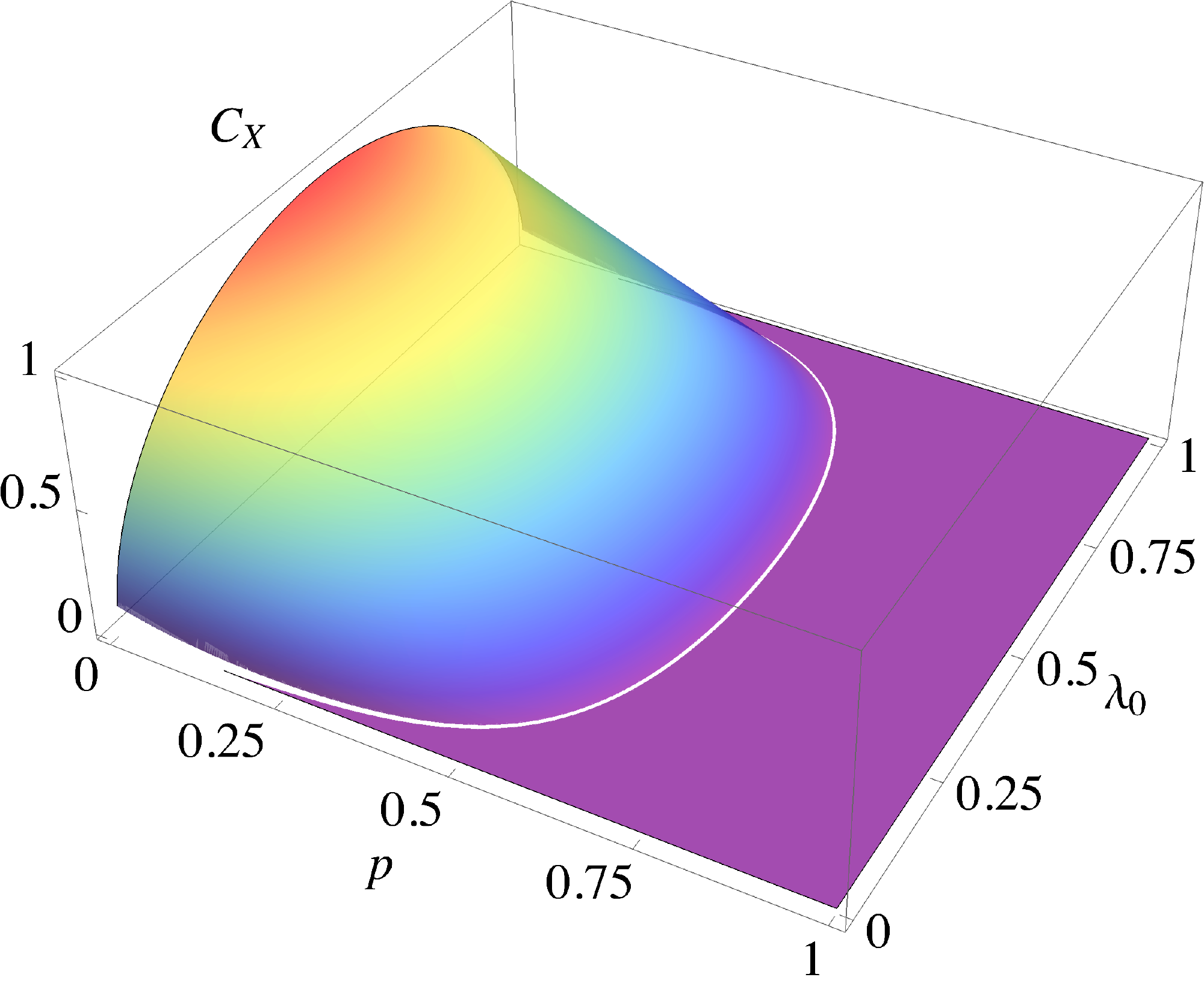}
\caption{Concurrence $C_X$ of the input state as a function of $p$ and $\lambda_0$. The white curve indicates the boundary between zero and positive concurrence.}
\label{fig1a}
\end{figure}

For this class of states the concurrence is given by,
\begin{align}
C_X=2~\text{max}[0, |\rho_{14}|-\sqrt{\rho_{22}\rho_{33}},|\rho_{23}|-\sqrt{\rho_{11}\rho_{44}}~],
\end{align}
which, in our case, reduces to
\begin{align}
C_X&=2~\text{max}[0, |\rho_{14}|-\sqrt{\rho_{22}\rho_{33}}~],\nonumber\\
&=2~\text{max}[0, (1-p)\sqrt{\lambda_0\lambda_1}-p/4] ,
\label{concrhox}
\end{align}
because $\rho_{23}=0$ and $\sqrt{\rho_{11}\rho_{44}}\geq 0$ and, therefore, $|\rho_{23}|-\sqrt{\rho_{11}\rho_{44}}\leq 0$. In what follows, we will use the expression for $C_X$ from Eq.~(\ref{concrhox}) for its simplicity. The region of positive concurrence of $\rho$ in the $(p,\lambda_0)$-plane is shown in Fig.~\ref{fig1a}.  The border between zero and positive concurrence is given by 
\begin{align}
p = 1- \frac{1}{1+4\sqrt{\lambda_0(1-\lambda_0)}} \,.
\label{lambdain}
\end{align}

We now ask for which values of the mixing parameter $p$ and the pure state entanglement parameter $\lambda_0$ do we get a positive concurrence of $\rho$, i.e., the domain for which $(1-p)\sqrt{\lambda_0\lambda_1}-p/4>0$. The allowed value of the mixing $p$ will be largest when $\ket{\psi}$ is maximally entangled. Therefore, setting $\lambda_0=\lambda_1=0.5$, we get the threshold condition, wherein $p<p_*=2/3$ for $\rho$ to be entangled \footnote{The state $\rho$ is a Werner state when $\ket{\psi}$ is a maximally entangled state. The threshold $p_*=2/3$ coincides with the distillation threshold of Werner states.}. Further, for any value of $p$ less than the threshold a nonzero concurrence is obtained only when $\lambda_0$ satisfies,
\begin{align}
\lambda_0\in (\frac{1}{2}-\Delta_0(p),\frac{1}{2}+\Delta_0(p)),
\label{lambdain}
\end{align}
where, $\Delta_0(p)=\frac{1}{2}\sqrt{1-\frac{p^2}{4(1-p)^2}}$. This implies that for a given value of $p<p_*$, the state $\ket{\psi}$ has to be sufficiently entangled for $\rho$ to be entangled since the entanglement of $\ket{\psi}$ decreases as one moves away from $\lambda_0=0.5$ in the domain $\lambda_0\in[0,1]$.

Next we want to find the average concurrence of the output state resulting from entanglement swapping two states of the form given in Eq.\ (\ref{noisyqubit}).  For the sake of simplicity, we shall assume that the states are the same and that we make standard Bell measurements in order to accomplish the entanglement swapping.  We start with the four-qubit state
\begin{equation}
\rho_{abcd} =  \left[ \frac{p}{4} I_{ab} + (1-p) |\psi\rangle_{ab}\langle\psi | \right] \left[ \frac{p}{4} I_{cd} + (1-p) |\psi\rangle_{cd}\langle\psi | \right]  ,
\label{proj_basis}
\end{equation}
and measure qubits $b$ and $c$ in the Bell basis.  We then need to find the average concurrence of the resulting state.

\begin{figure}
\centering
\includegraphics[width=\columnwidth]{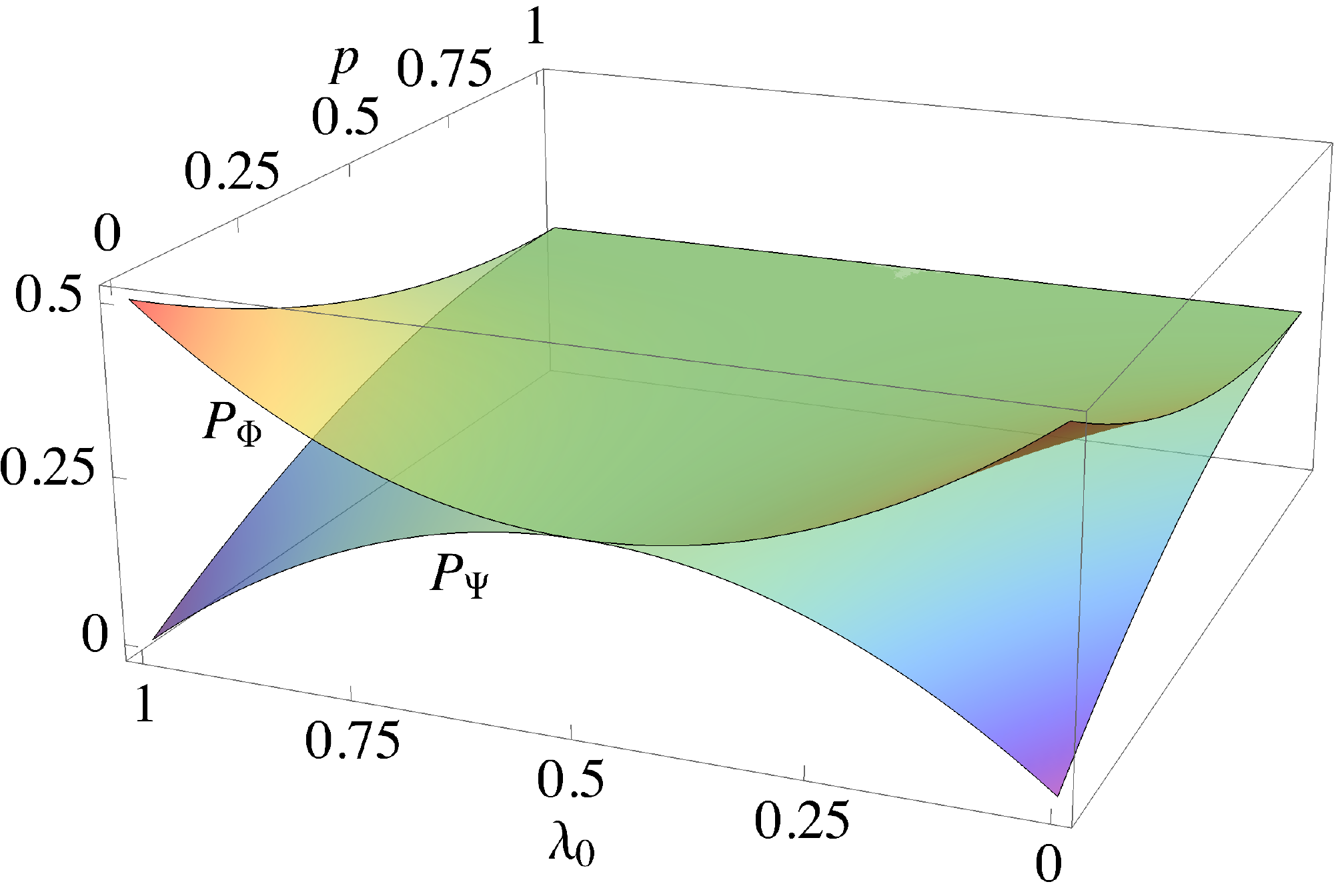}
\caption{The probabilities $P_{\Phi}$ and $P_{\Psi}$ as functions of $p$ and $\lambda_{0}$.}
\label{prob}
\end{figure}

\begin{figure}
\centering
\includegraphics[width=\columnwidth]{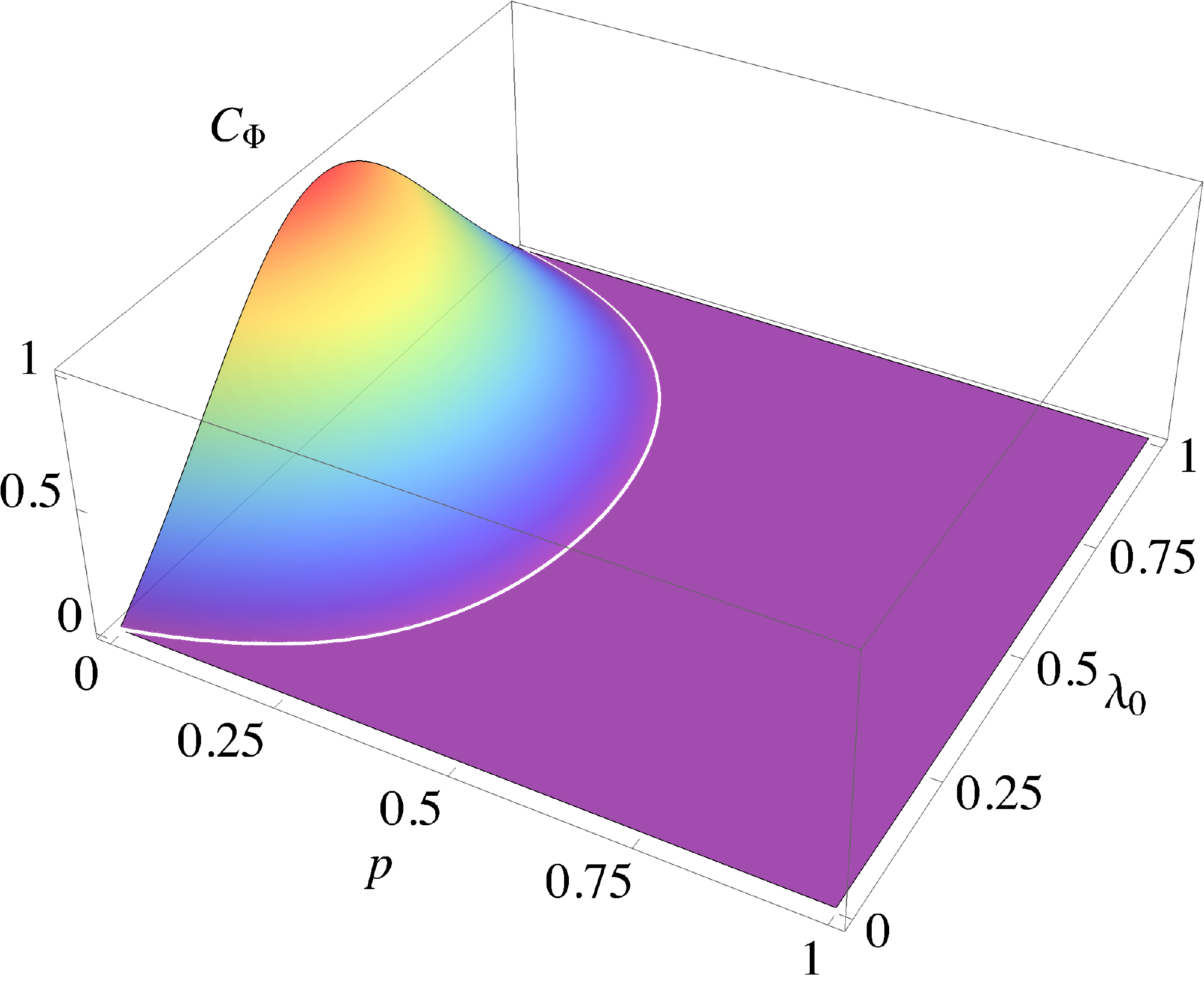}
\caption{Concurrence $C_\Phi$ in Eq~(\ref{concrhooutphi}) of the output state after the entanglement swapping  as a function of $p$ and $\lambda_0$. The white curve indicates the boundary between zero and positive concurrence.}
\label{C-Phi-out-3D}
\end{figure}

\begin{figure}
\centering
\includegraphics[width=\columnwidth]{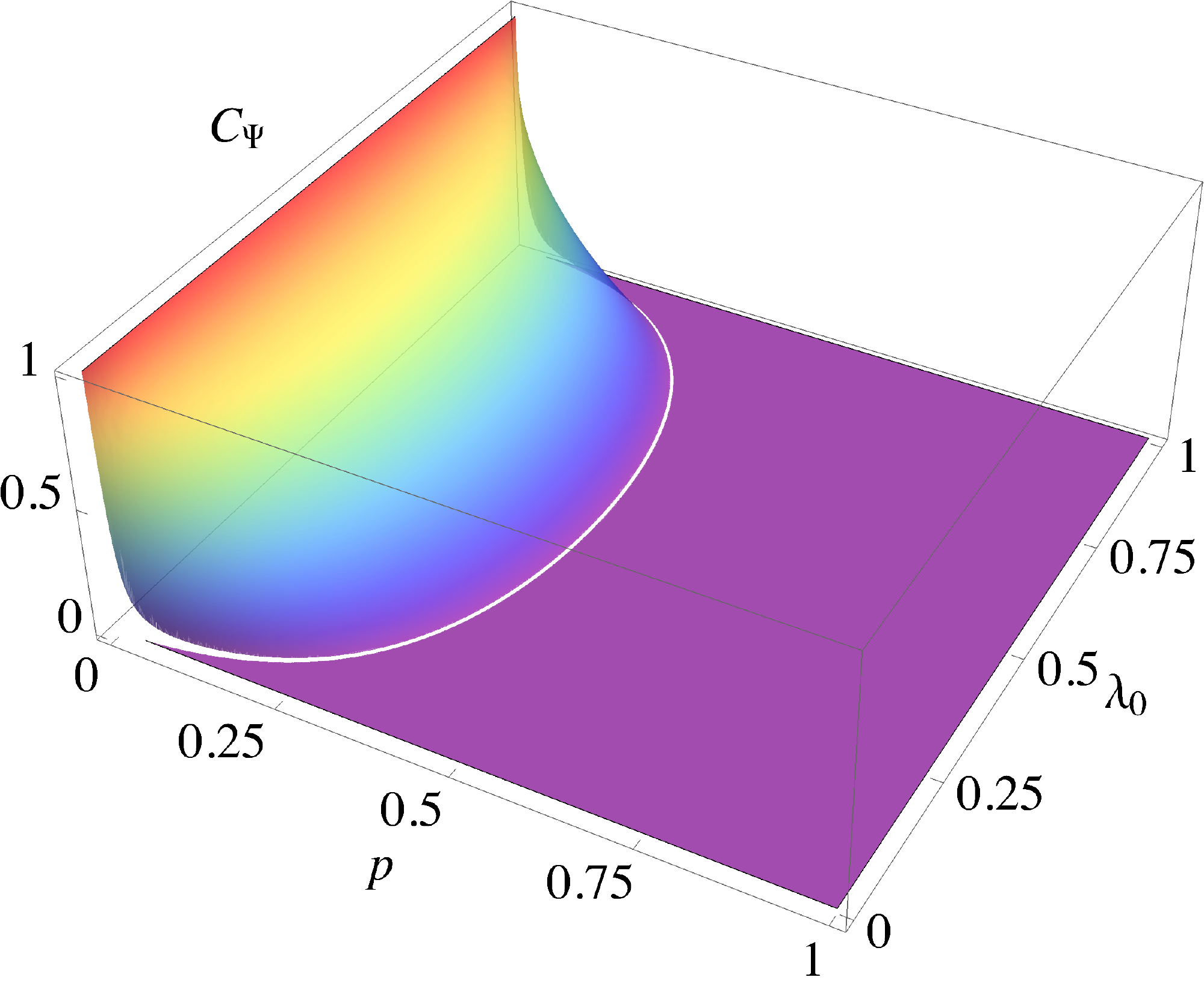}
\caption{Concurrence $C_\Psi$ in Eq~(\ref{concrhooutpsi}) of the output state after the entanglement swapping as a function of $p$ and $\lambda_0$. The white curve indicates the boundary between zero and positive concurrence.}
\label{C-Psi-out-3D}
\end{figure}

\begin{figure}
\centering
\includegraphics[width=\columnwidth]{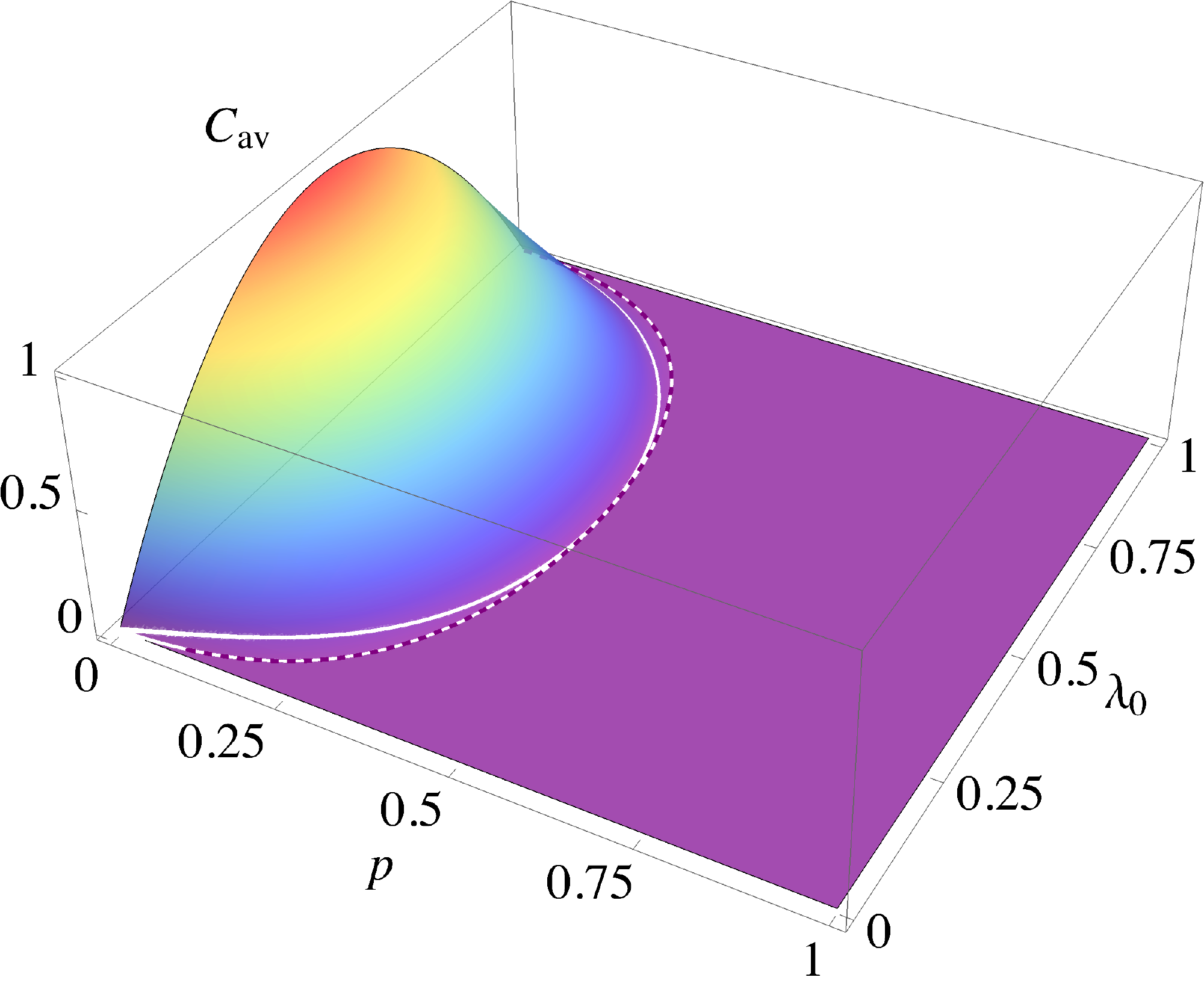}
\caption{Average concurrence $C_{av}$ of the output state after the entanglement swapping as a function of $p$ and $\lambda_0$. The boundary between zero and positive concurrence is indicated by the solid white curve for $C_\Phi$ and the dashed white curve for $C_\Psi$.}
\label{C-av-3D}
\end{figure}

The result of the measurement will be one of the four states
\begin{eqnarray}
|\Phi_{\pm}\rangle_{bc} & = & \frac{1}{\sqrt{2}} ( |00\rangle_{bc} \pm |11\rangle_{bc}) \nonumber \\
|\Psi_{\pm}\rangle_{bc} & = & \frac{1}{\sqrt{2}} ( |01\rangle_{bc} \pm |10\rangle_{bc}) .
\end{eqnarray}
If we obtain $|\Phi_{\pm}\rangle_{bc}$ the resulting state is
\begin{eqnarray}
\rho^{\Phi_{\pm}}_{ad} & = & \frac{1}{P_{\Phi }} \left[ D_{ad} + \frac{1}{2} (1-p)^{2} (\lambda_{0} |00\rangle_{ad} \pm \lambda_{1}|11\rangle_{ad})  \right. \nonumber \\
& & \left.  (\lambda_{0}\,_{ad}\langle 00| \pm \lambda_{1}\,_{ad}\langle 11|) \right] ,
\label{postmmtstatephi}
\end{eqnarray}
and if we obtain $|\Psi_{\pm}\rangle_{bc}$, the state is
\begin{eqnarray} 
\rho^{\Psi_{\pm}}_{ad} & = & \frac{1}{P_{\Psi}} \left[ D_{ad} + \frac{1}{2} (1-p)^{2} \lambda_{0}\lambda_{1} ( |01\rangle_{ad} \pm |10\rangle_{ad}) \right.  \nonumber \\
& & \left. (\,_{ad}\langle 10| \pm \,_{ad}\langle 10| ) \right] .
 \label{postmmtstatepsi}
\end{eqnarray}
Here,
\begin{eqnarray}
D_{ad} & = & \frac{p^{2}}{16} I_{ad} + (1-p)\frac{p}{8} [ (\lambda_{0} |0\rangle_{a}\langle 0| + \lambda_{1} |1\rangle_{a}\langle 1| )\otimes I_{d} \nonumber \\
& & + I_{a} \otimes (\lambda_{0} |0\rangle_{d}\langle 0| + \lambda_{1} |1\rangle_{d}\langle 1| ) ]  .
\end{eqnarray}
The probabilities of obtaining either $|\Phi_{+}\rangle_{bc}$ or $|\Phi_{-}\rangle_{bc}$ as a result of the Bell state measurement are the same,  and we denote them by $P_{\Phi}$.  Similarly, the probabilities of obtaining $|\Psi_{+}\rangle_{bc}$ and $|\Psi_{-}\rangle_{bc}$ are the same, and we denote them by $P_{\Psi}$.  Explicitly, we find (see Fig.~\ref{prob}),
\begin{align}
P_{\Phi}&=\frac{p(2-p)}{4}+ \frac{1}{2}  (1-p)^2 (\lambda_0^2+\lambda_1^2),\nonumber\\
P_{\Psi}&=\frac{p(2-p)}{4}+(1-p)^2\lambda_0\lambda_1.
\end{align}
In order to evaluate the average concurrence of the output state, we find the concurrence of each of the four density matrices above, and then weight them with the appropriate probability. The concurrence of the states $\rho_{ad}$ given in Eqs.~(\ref{postmmtstatephi}) and (\ref{postmmtstatepsi}) for any of the four measurement outcomes above can be obtained by noticing that they are also of the X-state form \cite{kirby}. Their explicit expressions are given by
\begin{align}
&C_X(\rho_{ad}^{\Phi_+})=C_X(\rho_{ad}^{\Phi_-})\nonumber\\
&=\frac{1}{P_{\Phi}}~\text{max}\left[ 0,(1-p)^2\lambda_0\lambda_1-\frac{p(2-p)}{8} \right] ,
\label{concrhooutphi}
\end{align}
and
\begin{align}
&C_X(\rho_{ad}^{\Psi_+})=C_X(\rho_{ad}^{\Psi_-})\nonumber\\
&=\frac{1}{P_{\Psi}}~\text{max} \left[ 0,(1-p)^2\lambda_0\lambda_1 \ \nonumber  \right.  \\
& \left. -\frac{p}{8}\sqrt{p^2+4p(1-p)+16(1-p)^2\lambda_0\lambda_1} \right]
\label{concrhooutpsi}
\end{align}
These are shown in Figs.\ \ref{C-Phi-out-3D} and \ref{C-Psi-out-3D}. The average concurrence of the entanglement-swapped state is then given by 
\begin{equation}
C_{av} = 2P_{\Phi} C_X(\rho_{ad}^{\Phi_+}) + 2P_{\Psi} C_X(\rho_{ad}^{\Psi_+}) .
\label{avconc}
\end{equation}
This is shown in Fig.\ \ref{C-av-3D}.

The threshold of $p$ for which the average concurrence, $C_{av}$, can be nonzero is obtained by numerically determining the thresholds for $C_X(\rho_{ad}^{\Phi_+})$ and $C_X(\rho_{ad}^{\Psi_+})$.  For both, we find the threshold $p_{*^,1}=(1-1/\sqrt{3})=0.422$. However, the threshold boundaries for these concurrences do not overlap in the $(p,\lambda_0)$ plane. This is because, for $p<p_{*,1}$, the state $\rho^{\Phi_\pm}_{ad}$ is entangled if
\begin{align}
\lambda_0\in(\frac{1}{2}-\Delta^{\Phi}_1(p),\frac{1}{2}+\Delta^{\Phi}_1(p)),
\end{align}
where $\Delta^{\Phi}_1(p)=\frac{1}{2}\sqrt{1-\frac{p(2-p)}{2(1-p)^2}}$, while, also for $p<p_{*,1}$, the state $\rho^{\Psi_\pm}_{ad}$ is entangled if,
\begin{align}
\lambda_0\in(\frac{1}{2}-\Delta^{\Psi}_1(p),\frac{1}{2}+\Delta^{\Psi}_1(p)),
\end{align}
where $\Delta^{\Psi}_1(p)=\frac{1}{2}\sqrt{1-\frac{p^2+p\sqrt{2p(2-p)}}{2(1-p)^2}}$. Since $\Delta^{\Psi}_1(p)>\Delta^{\Phi}_1(p)$ for $p<p_{*,1}$, we find that the swapped states $\rho^{\Psi_\pm}_{ad}$ may be entangled over a larger region of $\lambda_0$ than $\rho^{\Phi_\pm}_{ad}$. Further, from Eq.~(\ref{lambdain}), we find that $\Delta_0(p)>\Delta^{\Psi}_1(p),\Delta^{\Phi}_1(p)$, therefore the region around $\lambda_0=1/2$ for nonzero concurrence becomes narrower for a given $p$ after an entanglement swap. 


\begin{figure}
\centering
\includegraphics[width=\columnwidth]{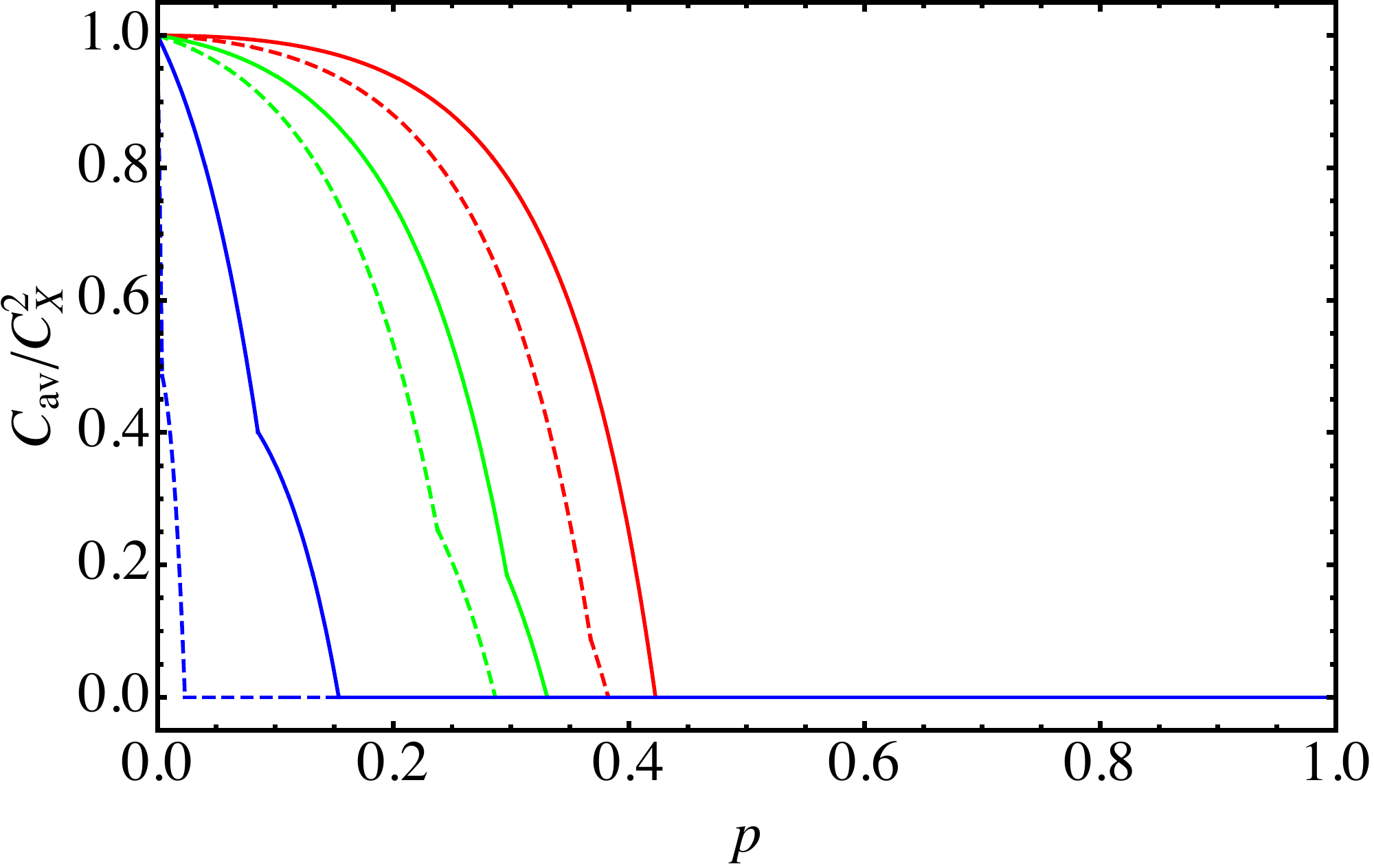}
\caption{The ratio $C_{av}/C_{X}^{2}$ as a function of $p$ for several values of $\lambda_{0}$. From left to right, the values of $\lambda_{0}$ are $0.01$ (dotted blue), $0.025$ (solid blue), $0.1$ (dotted green), $0.15$ (solid green), $0.25$ (dotted red), and $0.5$ (solid red).}
\label{Ratio}
\end{figure}


For pure states, the ratio $C_{av}/C_{X}^{2}$ is one, and we now have the results at hand to see how it behaves for mixed states of the form given in Eq.\ (\ref{noisyqubit}).  The ratio is plotted as a function of $p$ for a number of different values of $\lambda_{0}$ in Fig.\ \ref{Ratio}.  As we can see, it is a decreasing function of $p$, and the smaller the entanglement in the original states (the smaller the $\lambda_{0}$), the faster the fall-off with $p$.  These results strongly suggest that for mixed states, the product of the concurrences of the initial states is an upper bound to the average concurrence of the entanglement swapped state.

\subsubsection{ An upper bound to the average concurrence}
It can be useful to have an upper bound to the average concurrence that can be obtained from the entanglement swapping of two mixed states.  Let us use a pure state decomposition for the two-qubit states $\rho^{ab}_1$, on $\mathcal{H}_{a}\otimes\mathcal{H}_{b}$  and $\rho^{cd}_2$ on $\mathcal{H}_{c}\otimes\mathcal{H}_{d}$. So we have,
\begin{align}
\rho^{ab}_1&=\sum_{i}p_i\ket{\psi^{ab}_i}\bra{\psi^{ab}_i},\nonumber\\
\rho^{cd}_2&=\sum_{j}q_j\ket{\phi^{cd}_j}\bra{\phi^{cd}_j} .
\label{pure_decomp}
\end{align}

For the pure state pair $\ket{\psi^{ab}_i},\ket{\phi^{cd}_j}$ we define,
\begin{align}
p_{ij(k)}&={\rm Tr}_{ad}\braket{\chi^{bc}_k\ket{\psi^{ab}_i\phi^{cd}_j}\bra{\psi^{ab}_i\phi^{cd}_j}\chi^{bc}_k},\nonumber\\
\rho^{ad}_{ij(k)}&=\frac{\braket{\chi^{bc}_k\ket{\psi^{ab}_i\phi^{cd}_j}\bra{\psi^{ab}_i\phi^{cd}_j}\chi^{bc}_k}}{p_{ij(k)}},
\end{align}
where $|\chi_{jk}\rangle$ is an orthonormal  basis of $\mathcal{H}_{b}\otimes\mathcal{H}_{b}$.   We then have that,
\begin{align}
C_{av}(\ket{\psi_i^{ab}},\ket{\phi_j^{cd}})=\sum_{k=1}^{d^2}p_{ij(k)}C(\rho^{ad}_{ij(k)})
\label{avconc_state_pair}
\end{align}
and
\begin{align}
r_k&={\rm Tr}_{ad}\braket{\chi^{bc}_k|\rho^{ab}_1\otimes\rho^{cd}_2|\chi^{bc}_k}=\sum_{ij}p_iq_jp_{ij(k)},\nonumber\\
\rho^{ad}_k&=\frac{\braket{\chi^{bc}_k|\rho^{ab}_1\otimes\rho^{cd}_2|\chi^{bc}_k}}{r_k} \nonumber\\
&=\frac{\sum_{mn}p_mq_np_{mn(k)}\rho^{ad}_{mn(k)}}{\sum_{uv}p_uq_vp_{uv(k)}}.
\end{align}
The average concurrence in this case is,
\begin{align}
\bar{C}&=\sum_{k=1}^{d^2}r_kC(\rho_k^{ad})=\sum_{k=1}^{d^2}\sum_{ij}p_iq_jp_{ij(k)}C(\rho_k^{ad})\nonumber\\
&=\sum_{k=1}^{d^2}\sum_{ij}p_iq_jp_{ij(k)}C(\frac{\sum_{mn}p_mq_np_{mn(k)}\rho^{ad}_{mn(k)}}{\sum_{uv}p_uq_vp_{uv(k)}})\nonumber\\
&\leq\sum_{k=1}^{d^2}\sum_{ij}p_iq_jp_{ij(k)}\frac{\sum_{mn}p_mq_np_{mn(k)}C(\rho^{ad}_{mn(k)})}{\sum_{uv}p_uq_vp_{uv(k)}}\nonumber\\
&=\sum_{mn}p_mq_n\sum_{k=1}^{d^2}p_{mn(k)}C(\rho^{ad}_{mn(k)})\nonumber\\
&=\sum_{mn}p_mq_nC_{av}(\ket{\psi_m^{ab}},\ket{\phi_n^{cd}}) .
\end{align}
To obtain this inequality we have used the convexity of the concurrence and, in the last line, Eq.~(\ref{avconc_state_pair}). 

Now we note from Eq.~(\ref{avconc_maxbasis}) that for any $m,n$ the average concurrence $C_{av}(\ket{\psi_m^{ab}},\ket{\phi_n^{cd}})$ is maximized for the basis $\ket{\chi^{bc}_k}$ that is maximally entangled (at least for qubits). Therefore, $C_{av}(\ket{\psi_m^{ab}},\ket{\phi_n^{cd}})$ is maximized by performing the swapping measurement in maximally entangled basis. We note that the tightness of the upper bound,
\begin{align}
\bar{C}\leq\sum_{mn}p_mq_nC_{av}(\ket{\psi_m^{ab}},\ket{\phi_n^{cd}}),
\label{avconc_bound}
\end{align}
depends on the optimality of the pure state decompositions in Eq.~(\ref{pure_decomp}). If the pure state decompositions are chosen to be the ones that minimize the concurrence then the bound above can be tight. For example, consider $\rho^{ab}_1=I_{ab}$ and $\rho^{cd}_2=I_{cd}$ and the following two pure state decompositions of the identity,
\begin{align}
I&=\frac{1}{4}(\ket{\Psi^+}\bra{\Psi^+}+\ket{\Psi^-}\bra{\Psi^-}+\ket{\Phi^+}\bra{\Phi^+}+\ket{\Psi^-}\bra{\Psi^-})\nonumber\\
I&=\frac{1}{4}(\ket{00}\bra{00}+\ket{11}\bra{11}+\ket{22}\bra{22}+\ket{33}\bra{33}).
\end{align}
In the first case, Eq.~(\ref{avconc_bound}) suggests $\bar{C}\leq1$, while in the second case $\bar{C}\leq0$. 

We can apply this reasoning to obtain a simple upper bound for the average concurrence of the entanglement swapped qubits resulting from the four-qubit density matrix in Eq.\ (\ref{proj_basis}).  First, note that $D_{ad}$ is an incoherent superposition of product states, so its concurrence is zero.  This implies that
\begin{eqnarray}
C(\rho_{ad}^{\Phi_{\pm}}) & \leq & \frac{1}{P_{\Phi}} (1-p)^{2}\sqrt{\lambda_{0}\lambda_{1}}   \nonumber \\
C(\rho_{ad}^{\Psi_{\pm}}) & \leq & \frac{1}{P_{\Psi}} (1-p)^{2}\sqrt{\lambda_{0}\lambda_{1}} ,
\end{eqnarray}
with the result that $C_{av} \leq 4(1-p)^{2} \lambda_{0}\lambda_{1}$.

\section{Qudits}

\subsection{Pure states}

We can also look at entanglement swapping for higher dimensional systems \cite{hardy,bouda}.  We will start with two states, the first in $\mathcal{H}_{a} \otimes \mathcal{H}_{b}$ and the second in $\mathcal{H}_{c}\otimes \mathcal{H}_{d}$, with all spaces having dimension $N$,
\begin{eqnarray}
|\Psi_{ab}\rangle & = & \sum_{j=0}^{N-1} \sqrt{\lambda_{j}} |u_{j}\rangle_{a} |v_{j}\rangle_{b} \nonumber \\
|\Psi_{cd}\rangle & = & \sum_{j=0}^{N-1} \sqrt{\lambda_{j}^{\prime}} |u_{j}^{\prime}\rangle_{a} |v_{j}^{\prime}\rangle_{b}  ,
\end{eqnarray}
where we have expressed both states in their Schmidt representations.  In order to quantify the entanglement of these states, we shall use the I-concurrence \cite{buzek}, which, for $|\Psi_{ab}\rangle$, is
\begin{equation}
C_{I}(\Psi_{ab}) = \sqrt{2 [1-{\rm Tr}(\rho_{a}^{2}) ]} = \left[ 2( 1 - \sum_{j=0}^{N-1} \lambda_{j}^{2} ) \right]^{1/2} ,
\end{equation}  
with a similar expression for $C_{I}(\Psi_{cd})$.  In the above expression, $\rho_{a}$ is the reduced density matrix of $|\Psi_{ab}\rangle$.

In order to accomplish the entanglement swapping we measure the $b$ and $c$ qudits in the basis
\begin{equation}
|\chi_{mn}\rangle_{bc} = \frac{1}{\sqrt{N}} \sum_{j=0}^{N-1} e^{2\pi i jm/N} |v_{j}\rangle_{b} |u_{j+n}^{\prime}\rangle_{c} .
\end{equation}
If the measurement yields the state $|\chi_{mn}\rangle_{bc}$, then the resulting state in $\mathcal{H}_{a} \otimes \mathcal{H}_{d}$ is
\begin{equation}
|\Psi_{ad}\rangle = \frac{\eta_{mn}}{\sqrt{N}} \sum_{j=0}^{N-1} \sqrt{\lambda_{j} \lambda_{j+n}^{\prime}} e^{-2\pi i jm/N}  |u_{j}\rangle_{a} |v_{j+n}^{\prime}\rangle_{d} .
\end{equation}
Here $\eta_{mn}$ is a normalization factor satisfying
\begin{equation}
\eta_{mn} ^{2} \frac{1}{N} \left( \sum_{j=0}^{N-1} \lambda_{j}\lambda_{j+n}^{\prime} \right) = 1,
\end{equation}
and the probability of obtaining the result $|\chi_{mn}\rangle_{bc}$ is $1/\eta_{mn}^{2}$.  The I-concurrence of the state $|\Psi_{mn}\rangle_{ad}$ is
\begin{equation}
C_{I}(\Psi_{mn}) = \left[ 2\left(1-(1/N^{2}) \sum_{j=0}^{N-1} \eta_{mn}^{4} (\lambda_{j}\lambda_{j+n}^{\prime} )^{2} \right) \right]^{1/2} .
\end{equation}
To obtain the average I-concurrence we multiply the above expression by the probability to obtain $|\chi_{mn}\rangle_{bc}$ and add the results, yielding
\begin{equation}
\label{Cavqudit}
C_{I}^{(av)} = \sqrt{2} \sum_{n=0}^{N-1} \left[ \left( \sum_{j=0}^{N-1} \lambda_{j}\lambda_{j+n}^{\prime} \right) ^{2} - \sum_{j=0}^{N-1} (\lambda_{j}\lambda_{j+n}^{\prime})^{2} \right]^{1/2} .
\end{equation}
 
 A simple application of this result is to the case in which the $ab$ state is general, but the $cd$ state is maximally entangled, that is, $\lambda_{j}^{\prime} = 1/N$ for all $j$.  We then find that
 \begin{equation}
 C_{I}^{(av)} = C_{I}(\Psi_{ab}) ,
 \end{equation}
 so that the I-concurrence of the output state is completely determined by that of $|\Psi_{ab}\rangle$.

We can also look at the case in which one of the initial states have a small I-concurrence and see what kind of limits this places on the average I-concurrence of the state after the entanglement swap.  Let
\begin{equation}
\epsilon = 1 - \sum_{j=0}^{N-1} \lambda_{j}^{2}  ,
\end{equation}
which implies that $C_{I}(\Psi_{ab}) = \sqrt{2\epsilon}$, and we shall assume that $\epsilon$ is small, in particular, that $\epsilon < 2(N-1)/N^{2}$.    We can assume, without loss of generality, that $\lambda_{0}$ is the largest of the $\lambda_{j}$, that is $\lambda_{j}\leq \lambda_{0}$, for $j>0$.  Now
\begin{equation}
\lambda_{0}^{2} + \sum_{j=1}^{N-1} \lambda_{j}^{2} \leq \lambda_{0}^{2} + (1- \lambda_{0} )^{2} ,
\end{equation} 
where we have used the fact that if $\sum_{j=1}^{N-1} \lambda_{j}^{2} \leq (\sum_{j=1}^{N-1} \lambda_{j})^{2}$.  From this we have
\begin{equation}
2\lambda_{0}^{2} -3\lambda_{0} + \epsilon \geq 0  .
\end{equation}
This implies that either 
\begin{equation}
0 \leq \lambda_{0} \leq \frac{1}{2} - \frac{1}{2}(1-2\epsilon )^{1/2} ,
\end{equation}
or 
\begin{equation}
\frac{1}{2} + \frac{1}{2}(1-2\epsilon )^{1/2} \leq \lambda_{0} \leq 1 .
\end{equation}
We want to show that $\lambda_{0}$ must be in the upper range.  Let us assume it is in the lower range and show that this violates our assumptions.  If $\lambda_{0}$ is in the lower range, then
\begin{equation}
\sum_{j=1}^{N-1} \lambda_{j} = 1-\lambda_{0} \geq \frac{1}{2} + \frac{1}{2}(1-2\epsilon )^{1/2}  .
\end{equation}
Let $m$ be the maximum value of $\lambda{_j}$ for $j>0$.  The minimum value of $m$ occurs when all of the $\lambda_{j}$ for $j>0$ are the same, and this implies that
\begin{equation}
m \geq \frac{1-\lambda_{0}}{N-1} \geq \frac{1}{N-1}\left[ \frac{1}{2} + \frac{1}{2}(1-2\epsilon )^{1/2} \right] .
\end{equation}
Now, we have assumed that $\lambda_{0} \geq m$.  We will not be able to satisfy this condition for $\lambda_{0}$ in the lower range if 
\begin{equation}
\frac{1}{N-1}\left[ \frac{1}{2} + \frac{1}{2}(1-2\epsilon )^{1/2} \right] > \frac{1}{2} - \frac{1}{2}(1-2\epsilon )^{1/2} .
\end{equation}
If this condition is satisfied, then $\lambda_{0}$ must be in the upper range.  The above condition is satisfied when $\epsilon < 2(N-1)/N^{2}$, which we have assumed to be true.  Therefore, $\lambda_{0}$ is in the upper range.  

We can now return to Eq.\ (\ref{Cavqudit}).  First, 
\begin{eqnarray}
\sum_{j=0}^{N-1} \lambda_{j}\lambda_{j+n}^{\prime} & = & \lambda_{0} \lambda_{n}^{\prime} + (1-\lambda_{0}) \nonumber \\
 & \leq & \lambda_{0}\lambda_{n}^{\prime} + \Delta 
\end{eqnarray}
where $\Delta = (1/2)- (1/2)(1-2\epsilon )^{1/2} \sim \epsilon /2$.  Next,
\begin{equation}
\sum_{j=0}^{N-1} (\lambda_{j}\lambda_{j+n}^{\prime})^{2} \geq (\lambda_{0}\lambda_{n}^{\prime})^{2} .
\end{equation}
These estimates give us that
\begin{eqnarray}
C_{I}^{(av)} & \leq & \sqrt{2} \sum_{n=0}^{N-1} [ 2\lambda_{0}\lambda_{n}^{\prime} \Delta + \Delta^{2} ]^{1/2} \nonumber \\
 & \leq & \sqrt{2} \sum_{n=0}^{N-1} [ (2\lambda_{0}\lambda_{n}^{\prime} \Delta )^{1/2}  + \Delta ] .
\end{eqnarray}
The Schwarz inequality gives us that
\begin{equation}
\sum_{n=0}^{N-1} (\lambda_{n}^{\prime} )^{1/2} \leq \sqrt{N} ,
\end{equation} 
so that
\begin{equation}
C_{I}^{(av)} \leq 2 (N\Delta)^{1/2} + N\Delta .
\end{equation}
From this inequality, we see that the important quantity is $N\Delta \sim NC_{I}(\Psi_{ab})^{2}/4$.  For $C_{I}(\Psi_{ab})$ of order $N^{-1/2}$, the bound is of order one, but if it is of order $1/N$, then the average I-concurrence is small.  This gives us an estimate of how small the I-concurrence of a state has to be before it can make the average concurrence of an entanglement swapped state small as well.

A simple example can illuminate the relation between the I-concurrences of the initial states and that of the final entanglement-swapped state.  We will assume that 
\begin{equation}
\lambda_{j} = \left\{ \begin{array}{cc} \frac{1}{M} & 0\leq j \leq M-1 \\ 0 & M\leq j \leq N-1 \end{array} \right. ,
\end{equation}
for $M<N/2$.  We choose $\lambda_{j}^{\prime}$ to be the same.  We then find that 
\begin{equation}
\sum_{j=0}^{N-1} \lambda_{j}\lambda_{j+n}^{\prime} = \sum_{j=0}^{M-1} \lambda_{j}\lambda_{j+n}^{\prime} = \frac{M-n}{M^{2}} ,
\end{equation} 
for $0\leq n \leq M-1$.  The sum is zero for $M\leq n \leq N-m$, and is $(n+M-N)/M^{2}$ for $N-M +1 \leq n \leq N-1$.   We then find that
\begin{eqnarray}
C_{I}^{(av)} & = & \frac{\sqrt{2}}{M^{2}} \left\{ \sum_{n=0}^{M-1} (M-n)(M-n-1)]^{1/2}  \right. \nonumber \\
& & + \left. \sum_{n=N-n+1}^{N-1} [(n+M-N)(n+M-N-1) ]^{1/2}  \right\} \nonumber \\
& = & \frac{\sqrt{2}}{M^{2}} \left\{ \sum_{k=1}^{M} [k(k-1)]^{1/2} + \sum_{k=1}^{M-1} [k(k-1)]^{1/2} \right\} .
\end{eqnarray} 
It is possible to obtain upper and lower bounds to this expression.  To obtain a lower bound just replace $k$ in the sums by $k-1$ yielding
\begin{equation}
C_{I}^{(av)} \geq \sqrt{2} \left(\frac{M-1}{M}\right)^{2} = \sqrt{2} \left( 1 - \frac{2}{M}\right) + O(1/M^{2}) .
\end{equation}
For an upper bound we can use the Schwarz inequality, for example, 
\begin{equation}
\sum_{k=1}^{M} [k(k-1)]^{1/2} \leq \frac{M}{2} (M^{2}-1)^{1/2} ,
\end{equation}
to yield
\begin{eqnarray}
C_{I}^{(av)} & \leq & \frac{M-1}{\sqrt{2}M} \left\{ \left(\frac{M+1}{M-1}\right)^{1/2} + \left(1-\frac{2}{M}\right)^{1/2} \right\} \nonumber \\
& = & \sqrt{2} \left( 1- \frac{1}{M} \right) + O(1/M^{2}) .
\end{eqnarray}
We also note that in this case
\begin{equation}
C_{I}(\Psi_{ab}) = \sqrt{2} \left( 1-\frac{1}{M}\right)^{1/2} .
\end{equation} 
What we see in this example is that, up to $O(1/M^{2})$, the square of the I-concurrence of one of the initial states times $1/\sqrt{2}$ coincides with an upper bound for the I-concurrence of the entanglement-swapped state, so the actual I-concurrence of the entanglement-swapped state is less than or equal to $1/\sqrt{2}$ times  the square of the I-concurrence of one of the initial states (up to $O(1/M^{2})$).  We also note that both bounds are increasing functions of $M$, which implies that increasing the number of basis states in the initial entangled states will lead to a greater average I-concurrence of the output state.

\subsection{Noisy qudits}
We will consider states that are a mixture of the maximally entangled state $\ket{\Phi^{(N)}}=\frac{1}{\sqrt{N}}\sum_{j=0}^{N-1}\ket{u_j}\ket{v_j}$ with the totally mixed state $I/N^2$ in $N\times N$ dimensions. Therefore, our noisy qudits are of the isotropic form,
\begin{align}
\rho^{(N)}(p)=p\frac{I}{N^2}+(1-p)\ket{\Phi^{(N)}}\bra{\Phi^{(N)}},
\label{instate_N}
\end{align}
which are a one-parameter family of mixed states that are invariant under twirling. The invariance property implies that, $\int dU U\otimes U^* \rho^{(N)}(p) U^\dagger\otimes U^{*\dagger}=\rho^{(N)}(p)$, where `$*$' denotes complex conjugation in a fixed basis and `$\dagger$' denotes the adjoint operation.

Entanglement swapping can be performed on two such states, $\rho_{ab}=\rho_{cd}=\rho^{(N)}(p)$, via a complete set of projective measurements $\ket{\chi_{mn}}\bra{\chi_{mn}}$ on the `$bc$' systems where the states $\ket{\chi_{mn}}$ are given in Eq. \ref{proj_basis}. We find that there are $N^2$ distinct output states $\rho_{ad}^{(mn)}$ corresponding to the values of $m,n\in\{0,1,...,N-1\}$ that each occur with probability $1/N^2$. These output states are given by,
\begin{align}
\rho_{ad}^{(mn)}&=p(2-p)\frac{I_{ad}}{N^2}+\frac{(1-p)^2}{N}(\sum_{k=0}^{N-1}e^{-2\pi i mk/N}\ket{u_{k-n}v'_{k}})\nonumber\\
&~~~\times(\sum_{l=0}^{N-1}e^{+2\pi i ml/N}\bra{u_{l-n}v'_{l}}).
\label{outstates_N}
\end{align}

Each of these output states is found to be related to the isotropic state $\rho^{(N)}(p')$, where $p'=p(2-p)$, via local unitaries $U_a(m,n)\otimes I_N$, with $U_a(m,n)=\sum_{r=0}^{N-1}\ket{u_{r-n}}\bra{u_r}e^{-2\pi i m r/N}$, by noticing that
\begin{align}
(U_a(m,n)\otimes I_d) \rho^{(mn)}_{ad} (U^\dagger_a(m,n)\otimes I_d)=\rho^{(N)}_{ad}(p').
\end{align}
Therefore, the I-concurrence of each $\rho^{(mn)}_{ad}$ is the same as that of  $\rho^{(N)}_{ad}(p')$ for all $m,n$. Entanglement swapping results in an average output concurrence of $C_I^{(av)}=C_I(p')$.

For mixed states the I-concurrence is the ensemble minimum over all pure state decompositions of $\rho^{(N)}(p)$,
\begin{align}
C_I(\rho):=\min_{p_i,\ket{\psi_i}}\{p_iC_I(\ket{\psi_i})~|~\sum_{i=1}^Np_i\ket{\psi_i}\bra{\psi_i}=\rho^{(N)}(p)\},
\end{align}
with $\sum_i p_i=1,p_i>0$. For isotropic states $\rho^{(N)}(p)$ the optimal decomposition is in the form of factorizable states for $F_p\leq1/N$ where $F_p:=\braket{\Phi|\rho^{(N)}(p)|\Phi}=1-p+p/N^2$ is the fidelity \cite{horodecki_isotropic}. For $F_p\in(1/N,1)$ we use the approach developed in Ref.~\cite{terhal_isotropic_eof} to find that the I-concurrence is given by the convex hull of the function,
\begin{align}
Q(F_p)=\min_{\bar{\mu}}\{K(\bar{\mu})|F_p=|\sum_{i=1}^N\sqrt{\mu_i}|^2/N; \sum_{i=1}^N\mu_i=1\},
\label{qfunc}
\end{align}
where $K(\bar{\mu})=\sqrt{1-\sum_{i}\mu_i^2}$. 

Using the method of Lagrange multipliers to solve the constrained minimization problem we obtain the cubic equation (with $x_i=\sqrt{\mu_i}$),
\begin{align}
4x_i^3-2\Lambda_1x_i-\Lambda_2=0,
\label{lag_mult}
\end{align}
which for fixed $\Lambda_1,\Lambda_2$ determines the whole set $\{x_i\}$. Note that for nonzero $\Lambda_2$ the above equation admits no solution that is zero. Further, since the sum of its roots, $\alpha+\beta+\gamma=0$, at most two of the roots can be positive. Therefore, all vectors $\bar{\mu}$ that are a solution to the minimization problem Eq.~(\ref{qfunc}) have entries $\mu_i\in\{\alpha^2,\beta^2\}$ where we denote the positive definite solutions of Eq. (\ref{lag_mult}) by $\alpha,\beta$. For any vector $\bar{\mu}$ with $r$ entries that are $\alpha^2$ and the rest $(N-r)$ entries that are $\beta^2$ the minimization problem is reduced to a minimization of,
\begin{align}
\sqrt{1-r\alpha^2-(N-r)\beta^2},
\label{min_func}
\end{align} 
for a fixed $r$. There are also the constraints which imply
\begin{align}
\{r\alpha^2+(N-r)\beta^2&=1, r\alpha+(N-r)\beta&=\sqrt{F_pN}\}.
\end{align}
The constraint equations can be solved to obtain
\begin{align}
\alpha^\pm_r(F_p)=\frac{\sqrt{rF_p}\pm\sqrt{(N-r)(1-F_p)}}{\sqrt{rN}},
\end{align}
and, correspondingly, $\beta^\pm_r(F_p)=(\sqrt{F_pN}-r\alpha^\pm(F_p))/(N-r)$. Since $\alpha^-_r=\beta^+_{N-r}$ and $\beta^-_r=\alpha^+_{N-r}$, the function in Eq. (\ref{min_func}) takes the same values for $\alpha^+_r$ and $\alpha^-_{N-r}$, therefore we restrict ourselves to the set $(\alpha,\beta)=(\alpha^+_r,\beta^+_r)$. The function $Q(F_p)$ is thus obtained as the point-wise minimum over possible choices of $r$ of the function,
\begin{align}
Q_r(F_p)=\sqrt{1-r\alpha^+_r(F_p)^4-(N-r)\beta^+_r(F_p)^4},
\label{min_func_m}
\end{align}
defined on the domain $r/N\leq F_p$. The restriction on the domain comes from requiring positivity of $\beta^+_r(F_p)$. One can verify that $\alpha^+_r(F_p)\geq\beta^+_r(F_p)$ for all $0\leq F_p\leq 1$ and $0\leq r\leq N$. The point-wise minimum of the function in Eq. (\ref{min_func_m}) is then obtained for $r=1$, i.e., $Q(F_p)=Q_1(F_p)$ is the desired function in Eq.~(\ref{qfunc}), for any $N\geq2$.

\begin{figure}
\centering
\includegraphics[width=\columnwidth]{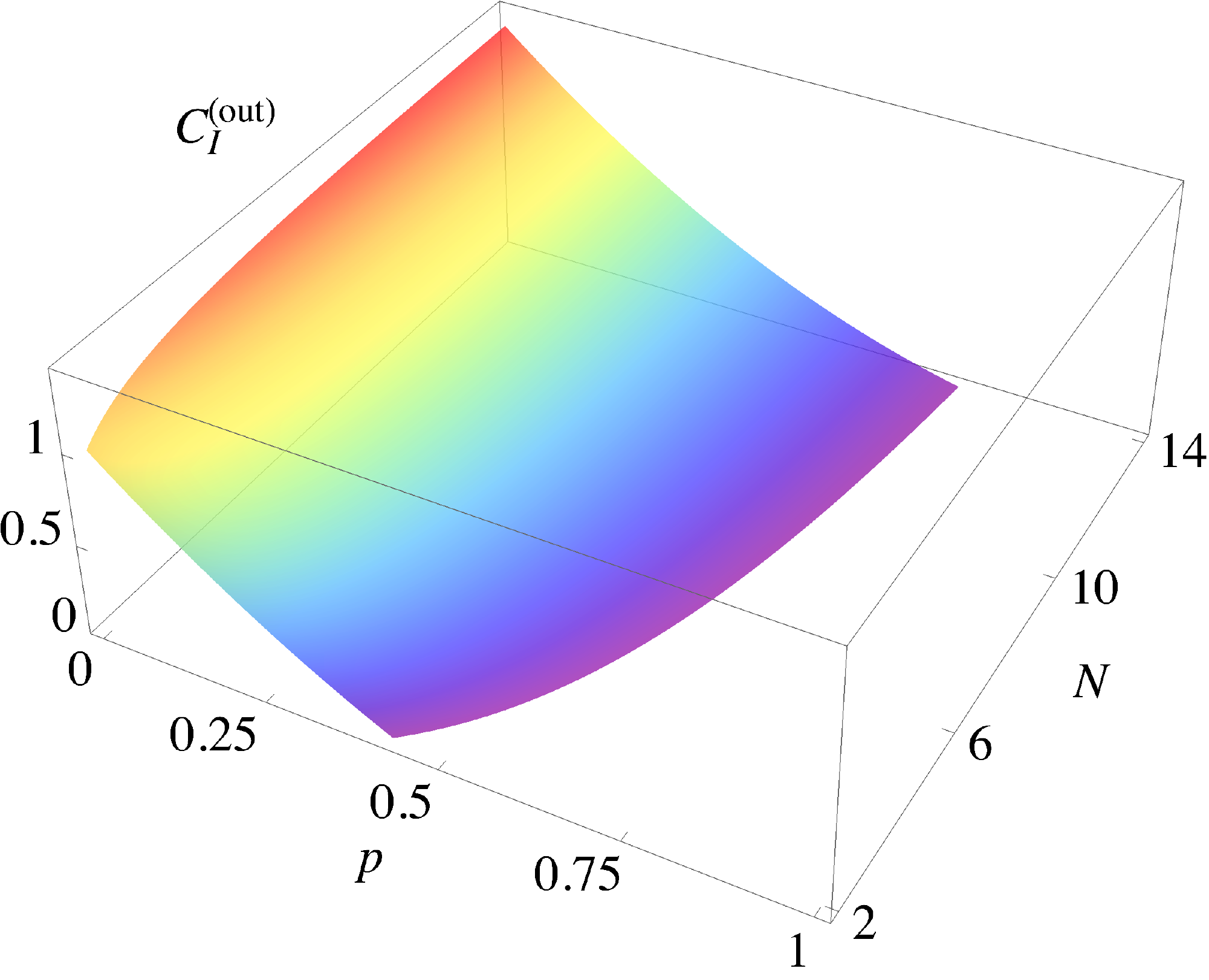}
\caption{I-concurrence of isotropic states vs.the mixing parameter $p$ in various dimensions $N$.}
\label{fig_Iconc}
\end{figure}

We find that the function $Q_1(F_p)$ is concave in the entire domain $F_p\in(1/N,1]\forall N\geq2$, therefore its convex hull is the straight line through the points $(1/N,0)$ and $(1,\sqrt{2(1-1/N)})$ in the $(F_p,C_I(\rho^{(N)}(p)))$-plane. Thus the I-concurrence of $\rho^{(N)}(p)$ is found to be (see Fig.\ \ref{fig_Iconc}),
\begin{align}
C_I(\rho^{(N)}(p)):=\begin{cases}0,~~~~~~~~~~~~~~~~~~~~~~~~~~~~~~~F_p\leq1/N,\\
\frac{\sqrt{2N}}{\sqrt{N-1}}F_p-\frac{\sqrt{2}}{\sqrt{N(N-1)}},~~~~1/N<F_p\leq 1.\end{cases}
\label{Iconc_final}
\end{align}

\begin{figure}
\centering
\includegraphics[width=\columnwidth]{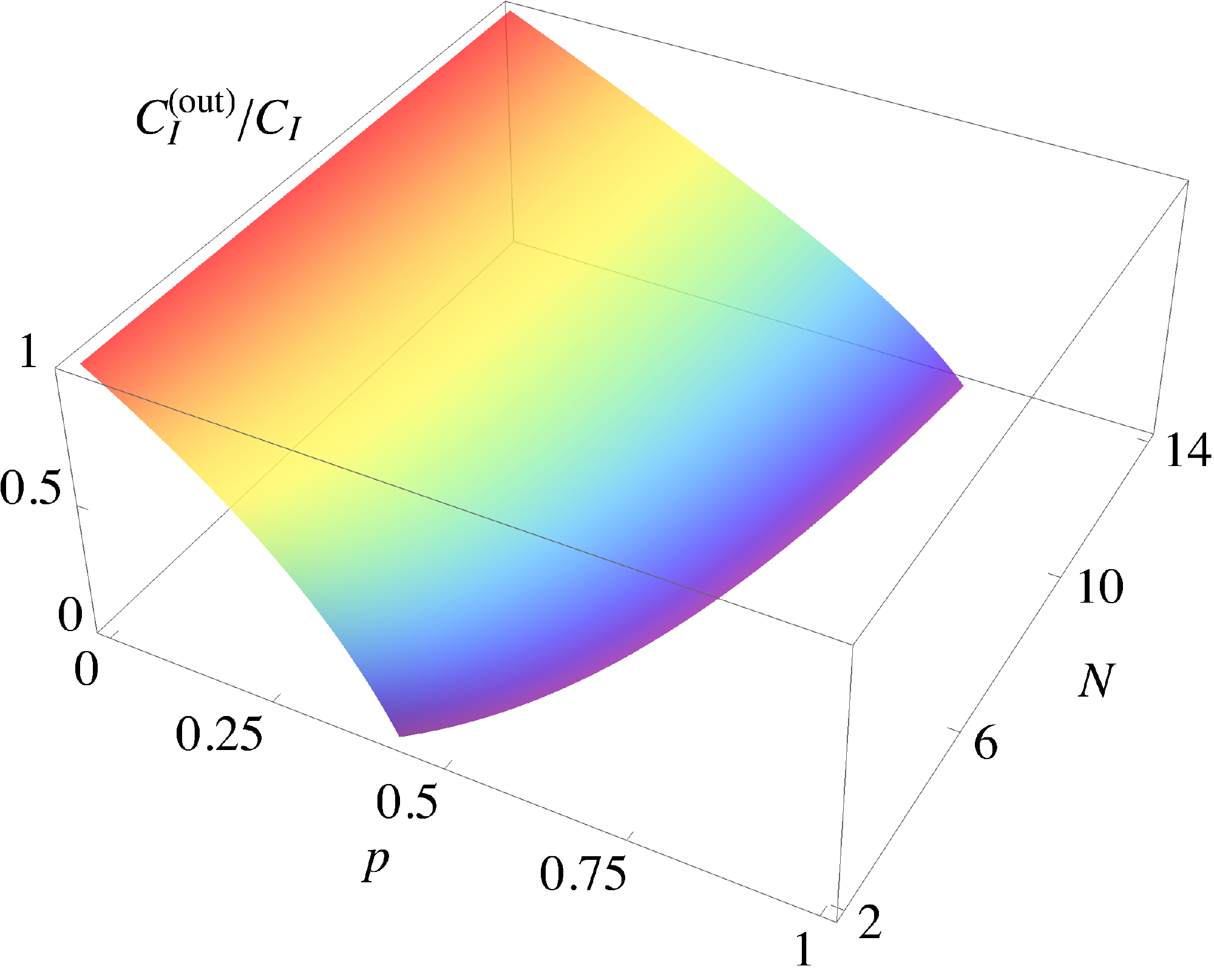}
\caption{Ratio of the I-concurrence of the output states to that of the input states for noisy qudits vs. the mixing parameter $p$ in various dimensions $N$.}
\label{fig_Iconc-ratio}
\end{figure}

The I-concurrence of each of the states $\rho^{(mn)}_{ad}$ in Eq.~(\ref{outstates_N}) is thus $C_I(\rho^{(N)}(p'))$ and equals the average I-concurrence.  The I-concurrence of the output states is shown as a function of $p$ and $N$ in Fig.\ \ref{fig_Iconc} and the ratio of the input I-concurrence to the output I-concurrence is shown in Fig.\ \ref{fig_Iconc-ratio}.

The threshold of the mixing parameter can now be obtained from Eq.~(\ref{Iconc_final}). For input states the threshold evaluates to $p<p_*=N/(N+1)$ whereas the threshold for the output states is given by $p_*=1-\sqrt{N^3-N^2-N+1}/(N^2-1)$.

\section{Conclusion}
We have studied the dependence of the average concurrence of the output state of entanglement swapping on the concurrences of the input states.  In the case of qubit pure states, the dependence is simple; the average concurrence of the output state is just the product of the concurrences of the input states.  This result is easily generalized to multiple swaps.  An example with mixed state inputs suggests that the product of the input concurrences is an upper bound for the output concurrence.  For qudits we made use of the I-concurrence.  While the relation between the I-concurrence of the input states and that of the output state is more complicated than in the qubit case, there are some cases in which the result is simple.  In particular, if one of the input states is maximally entangled, the output I-concurrence is the same as the I-concurrence of the non-maximally entangled input state.  Finally, we examined the entanglement swapping of noisy qudits for a particular class of qudits, and found how the average I-concurrence of the entanglement swapped states depends on that of the input qudits.

\emph{Acknowledgement.} Research of JB, MH, and DF was sponsored by the Army Research Laboratory and was accomplished under Cooperative Agreement Number W911NF-20-2-0097.


\begin{thebibliography}{99}
\bibitem{Zukowski} M.~Zukowski, A.~Zeilinger, M.~A.~Horne, and A.~K.~Ekert, \prl {\bf 71}, 4287 (1993).
\bibitem{Zeilinger} T.~Jennewein, G.~Weihs, J.-W.~Pan, and A.~Zeilinger, \prl {\bf 88}, 017903 (2001).
\bibitem{Gisin} H.~de~Riedmatten, I.~Marcikic, J.~A.~W.~van~Houwelingen, W.~Tittel, H.~Zbinden, and N.~Gisin, \pra {\bf 71}, 050302(R) (2005).
\bibitem{bose} S.~Bose, V.~Vedral, and P.~L.~Knight, \pra {\bf 57}, 822 (1998).
\bibitem{knight} S.~Bose, V.~Vedral, and P.~L.~Knight, \pra {\bf 60}, 194 (1999).
\bibitem{briegel} H.-J.~Briegel, W.~D\"{u}r, J.~I.~Cirac, and P.~Zoller, \prl {\bf 81}, 5932 (1998).
\bibitem{duan} L.-M.~Duan, M.~Lukin, J.~I.~Cirac, and P.~Zoller, Nature (London), {\bf 414}, 413 (2001).
\bibitem{razavi} M.~Razavi, M.~Piani, and N.~L\"{u}tkenhaus, \pra {\bf 80}, 032301 (2009).
\bibitem{lukin} S.~Muralidharan, L.~Li, J.~Kim, N.~L\"{u}tkenhaus, M.~D.~Lukin, and L.~Jiang, Sci.\ Rep.\ {\bf 6}, 20463 (2015).
\bibitem{network}  S.~Perseguers, J.~I.~Cirac, A.~Acin, M.~Lewenstein, and J.~Wehr, \pra {\bf 77}, 022308 (2008).
\bibitem{sen} A.~Sen(De), U.~Sen, \v{C}.~Brukner, V.~Bu\v{z}ek, and M.~Zukowski, \pra {\bf 72}, 042310 (2005).
\bibitem{wojcik} A.~Wojcik, J.~Modlawska, A.~Grudka, and M.~Czechlewski, Phys.\ Lett.\ A {\bf 374}, 4831 (2010).
\bibitem{klobus} W.~Klobus, W.~Laskowski, M.~Markiewicz, and A.~Grudka, \pra {\bf 86}, 020302 (2012).  
\bibitem{grudka} J.~Mod\l{}awska and A.~Grudka, \pra {\bf 78}, 032321 (2008).
\bibitem{roa} L.~Roa, A.~Mu\~{n}oz, and G.~Gr\"{u}ning, Phys.\ Rev.\ A {\bf 89}, 064301 (2014).
\bibitem{kirby} B.~T.~Kirby, S.~Santra, V.~S.~Malinovsky, and M.~Brodsky, \pra {\bf 94}, 012336 (2016).
\bibitem{hardy}L.~Hardy and D.~D.~Song, \pra {\bf 62}, 052315 (2000).
\bibitem{bouda} J.~Bouda and V.~Bu\v{z}ek, J.\ Phys.\ A {\bf 34}, 4301 (2001).
\bibitem{buzek}  P. Rungta, V.\ Bu\v{z}ek, C.~M.\ Caves, M.\ Hillery, and G.\ J.\
Milburn, \pra {\bf 64}, 042315 (2001).
\bibitem{xstate_yu} T.~Yu and J. H.~Eberly, Quantum\ Info. \ Comput. {\bf 7}, 459 (2007).  
\bibitem{horodecki_isotropic} M.~Horodecki and P. Horodecki, arXiv:quant-ph/9708015 (1997).
\bibitem{terhal_isotropic_eof} B.~M.~Terhal and K. G. H.~Vollbrecht, \prl {\bf 85}, 2625 (2000).
\bibitem{bacco} D.~Bacco, J.~F.~F.~Bulmer, M. Erhard, M.~Huber, and S.~Paesani, arXiv:2103.09202 .
\end{thebibliography}
\end{document}